\documentclass{JHEP3}

\usepackage{amsmath}
\usepackage{amsfonts}
\usepackage{amssymb}
\usepackage{graphicx}
\usepackage{citesort}
\usepackage{mathrsfs}

\def\S{{\mathbb S}}
\def\V{{\mathscr V}}

\newcommand{\alg}[1]{\mathfrak{#1}}
\newcommand{\su}{\alg{su}}

\def\ads{{\rm AdS}_5\times {\rm S}^5}

\author{Marius de Leeuw \footnote{E-mail: M.deLeeuw@uu.nl}
 \\  {\it Institute for Theoretical
Physics and Spinoza Institute,\\ Utrecht University, 3508 TD
Utrecht, The Netherlands}}

\abstract{We reformulate the nested coordinate Bethe ansatz in
terms of coproducts of Yangian symmetry generators. This allows us
to derive the nested Bethe equations for arbitrary bound state
string S-matrices. The bound state number dependence in the Bethe
equations appears through the parameters $x^{\pm}$ and the
dressing phase only.}

\title{The Bethe Ansatz for $\ads$ Bound States}

\preprint{
          \tiny{ITP-UU-08-52}\\[-.5ex]
          \tiny{SPIN-08-42}\\[-.5ex]
          }

\begin{document}

\section{Introduction and summary}

The AdS/CFT correspondence \cite{Maldacena:1997re} has been the
subject of intensive research. This duality provides a powerful
tool to study a plethora of interesting topics in theoretical
physics. One of the best studied examples is $\mathcal{N}=4$ SYM
gauge theory which is conjectured to be dual to superstring theory
on $\ads$. However, even testing or proving the duality for these
concrete models is a difficult problem.\smallskip

A breakthrough in the understanding of this duality was the
discovery of integrable structures. Integrability was found in
$\mathcal{N}=4$ super Yang-Mills theory by the appearance of
(integrable) spin chains describing the operator spectrum
\cite{Minahan:2002ve}. The classical string sigma model on $\ads$
was also shown to be integrable \cite{Bena:2003wd}. If the full
quantum theories also exhibit integrable structures, then that
would constrain them considerably. For example, in scattering
processes for integrable theories the set of particle momenta is
conserved and scattering processes always factorize into a
sequence of two-body interactions. This implies that in such
theories the scattering information is encoded in the two-body
S-matrix. Unfortunately, a full proof of integrability of both
theories is currently still lacking, but nevertheless there is a
lot of evidence hinting that integrability is a feature of the
full quantum theories.\smallskip

By assuming integrability one can make use of the S-matrix
approach. This proved to be a powerful instrument to study the
operator spectrum of $\mathcal{N}=4$ SYM.
\cite{Minahan:2002ve,Serban:2004jf,Beisert:2004hm,Staudacher:2004tk}.
This lead to the conjecture of the ``all-loop" Bethe equations
describing the gauge theory asymptotic spectrum
\cite{Beisert:2005fw,Beisert:2005tm}. For the $\ads$ superstring,
based on the knowledge of the classical finite-gap solutions
\cite{Kazakov:2004qf}, a Bethe ansatz for the $\su(2)$ sector was
proposed \cite{Arutyunov:2004vx}. Finally, in both cases, exact
two body S-matrices were found
\cite{Beisert:2005tm,Arutyunov:2006yd}. These enabled the use of
the Bethe ansatz
\cite{Beisert:2005tm,Martins:2007hb,Leeuw:2007uf}, confirming the
conjectured Bethe equations for physical states.\smallskip

More precisely, the two body S-matrix (scattering fundamental
multiplets) that appears in this approach, is almost completely
fixed by symmetry. Both the asymptotic spectrum of $\mathcal{N}=4$
super Yang-Mills theory \cite{Beisert:2004hm,Beisert:2005tm} as
well as the light-cone Hamiltonian
\cite{Arutyunov:2004yx,Arutyunov:2006ak,Frolov:2006cc} for the
$\ads$ superstring exhibit the same symmetry algebra; centrally
extended $\mathfrak{su}(2|2)$. The requirement that the S-matrix
is invariant under this algebra determines it uniquely up to an
overall phase factor \cite{Beisert:2005tm} and the choice of
representation basis \cite{Arutyunov:2006yd}. In a suitable local
scattering basis, the S-matrix respects most properties of massive
two-dimensional integrable field theory, like unitarity, crossing
symmetry and the Yang-Baxter equation \cite{Arutyunov:2006yd}.
Furthermore, the (dressing) phase appeared to be a striking
feature of the string S-matrix and it has been studied intensely,
see e.g.
\cite{Beisert:2006ib,Freyhult:2006vr,Eden:2006rx,Beisert:2007hz}.
By combining its expansion in terms of local conserved charges
with crossing symmetry \cite{Janik:2006dc}, one can find
interesting solutions \cite{Beisert:2006ib,Beisert:2006ez}, which
incorporates string and gauge theory data.\smallskip

Actually, apart from the multiplet of fundamental particles, the
string sigma model also contains an infinite number of bound
states \cite{Arutyunov:2007tc}. These bound states fall into short
(atypical) symmetric representations of the centrally extended
$\mathfrak{su}(2|2)$ algebra
\cite{Dorey:2006dq,Beisert:2006qh,Chen:2006gp}. Of course, these
states scatter via their own S-matrices. Recently a number of
these bound states S-matrices have been found; they describe
scattering processes involving fundamental multiplets and
two-particle bound state multiplets \cite{Arutyunov:2008} and the
scattering of a fundamental multiplet with an arbitrary bound
state multiplet \cite{Bajnok:2008bm}. However, invariance under
centrally extended $\mathfrak{su}(2|2)$ is not enough to fix all
these S-matrices. One needs to impose the Yang-Baxter equation by
hand in order to fix them up to a phase. The overall phase that
still remains can be chosen to satisfy crossing symmetry
\cite{Arutyunov:2008}.\smallskip

Both the fundamental and the two particle bound state S-matrices
were shown to exhibit a larger symmetry algebra \footnote{See
\cite{Arutyunov:2006yd} for an earlier discussion of higher
symmetries of the fundamental S-matrix.} of Yangian type
\cite{Beisert:2007ty,Matsumoto:2007rh,Beisert:2007ds,deLeeuw:2008dp}.
Moreover, as an alternative to the Yang-Baxter equation, the bound
state S-matrices appear to be completely fixed (again up to an
overall scale), by requiring invariance under Yangian symmetry
\cite{deLeeuw:2008dp}. This seems to indicate that the Yangian
symmetry is restrictive enough to fix all bound state S-matrices
and perhaps should be seen as the fundamental scattering
symmetry.\smallskip

The formulated asymptotic Bethe ans\"atze, following from the
S-matrix approach, only describe the spectra in the infinite
volume limit. However, for a full check of this particular case of
the AdS/CFT correspondence, the complete spectra have to be
computed and compared. Away from the asymptotic region, matters
become more involved because wrapping interactions appear. One way
to include these is L\"uscher's perturbative approach
\cite{Janik:2007wt,Gromov:2008ie,Heller:2008at,Ambjorn:2005wa}. In
L\"uscher's approach one deals with corrections coming from
virtual particles that propagate around the compact direction.
These virtual particles can be both fundamental particles and
bound states. This method has proven quite successful since it
recently allowed the computation of the full four-loop Konishi
operator, including wrapping interactions
\cite{Sieg:2005kd,Bajnok:2008bm}. The result coincided with the
gauge theory computation
\cite{Fiamberti:2007rj,Fiamberti:2008sh,Fiamberti:2008sm},
providing an extremely non-trivial check of the
correspondence.\smallskip

The approach by L\"uscher is closely related to the thermodynamic
Bethe ansatz (TBA) \cite{Zamolodchikov:1989cf}. In the TBA one
deals with finite size effects by defining a mirror model
\cite{Arutyunov:2007tc}. Finite size effects in the original
theory correspond to finite temperature effects in the infinite
volume for the mirror theory. Here again, one needs to include all
(physical) bound states. One of the advantages of this method is
that one can still use the asymptotic Bethe ansatz. So far, this
has only been carried out for the fundamental multiplets
\cite{Beisert:2005tm,Martins:2007hb,Leeuw:2007uf}.\smallskip

In other words, knowledge of the bound states, their S-matrices
and the corresponding Bethe ans\"atze is crucial for a complete
understanding of finite size effects. Because of the relation
between Yangian symmetry and the bound state S-matrices, one might
suspect that the Bethe equations are also closely related to
Yangian symmetry. This indeed appears to be the case as we will
explain in this note. The Bethe equations that are derived can be
used to study bound states of the mirror model by analytic
continuation \cite{Arutyunov:2007tc}.
\smallskip

The aim of this paper is to provide a rigorous derivation of the
Bethe Ansatz equations for the bound states, by diagonalizing the
multi-particle bound state S-matrices using Yangian symmetry as a
tool. The explicit bound state number dependence in the Bethe
equations appears through the parameters $x^{\pm}$ and the
dressing phase only. The Bethe equations emerging in this way,
were found, a posteriori, to follow from a fusion procedure. This
justifies the fusion procedure at the level of the Bethe ansatz
equations.
\smallskip

The paper is organized as follows. In the first section we recall
the basic formulas of the Yangian of the centrally extended
$\su(2|2)$ algebra in the superspace formalism. Subsequently, we
will discuss the Bethe ansatz for fundamental representations.
Finally, the Bethe ansatz will be reformulated in terms of
coproducts of the Yangian and the bound state Bethe equations will
be derived.

\section{Symmetry algebra, coproducts and S-matrices}

In this paper the full (Yangian) symmetry algebra of the S-matrix
is important. In this section we will give a brief overview of the
Yangian symmetry of the bound state S-matrices. For more details
see e.g \cite{deLeeuw:2008dp} and references therein.

\subsection{The algebra in superspace}

We will first discuss centrally extended $\mathfrak{su}(2|2)$.
This algebra is the symmetry algebra of the $\ads$ superstring and
it is also the symmetry algebra of the spin chain connected to
$\mathcal{N}=4$ SYM. The algebra has bosonic generators
$\mathbb{R},\mathbb{L}$, supersymmetry generators
$\mathbb{Q},\mathbb{G}$ and central charges
$\mathbb{H},\mathbb{C},\mathbb{C}^{\dag}$. The non-trivial
commutation relations between the generators are given by
\begin{eqnarray}
\begin{array}{lll}
\ [\mathbb{L}_{a}^{\ b},\mathbb{J}_{c}] = \delta_{c}^{b}\mathbb{J}_{a}-\frac{1}{2}\delta_{a}^{b}\mathbb{J}_{c} &\qquad & \ [\mathbb{R}_{\alpha}^{\ \beta},\mathbb{J}_{\gamma}] = \delta_{\gamma}^{\beta}\mathbb{J}_{\alpha}-\frac{1}{2}\delta_{\alpha}^{\beta}\mathbb{J}_{\gamma}\\
\ [\mathbb{L}_{a}^{\ b},\mathbb{J}^{c}] = -\delta_{a}^{c}\mathbb{J}^{b}+\frac{1}{2}\delta_{a}^{b}\mathbb{J}^{c} &\qquad& \ [\mathbb{R}_{\alpha}^{\ \beta},\mathbb{J}^{\gamma}] = -\delta^{\gamma}_{\alpha}\mathbb{J}^{\beta}+\frac{1}{2}\delta_{\alpha}^{\beta}\mathbb{J}^{\gamma}\\
\ \{\mathbb{Q}_{\alpha}^{\ a},\mathbb{Q}_{\beta}^{\
b}\}=\epsilon_{\alpha\beta}\epsilon^{ab}\mathbb{C}&\qquad&\ \{\mathbb{G}^{\ \alpha}_{a},\mathbb{G}^{\ \beta}_{b}\}=\epsilon^{\alpha\beta}\epsilon_{ab}\mathbb{C}^{\dag}\\
\ \{\mathbb{Q}_{\alpha}^{a},\mathbb{G}^{\beta}_{b}\} =
\delta_{b}^{a}\mathbb{R}_{\alpha}^{\ \beta} +
\delta_{\alpha}^{\beta}\mathbb{L}_{b}^{\ a}
+\frac{1}{2}\delta_{b}^{a}\delta_{\alpha}^{\beta}\mathbb{H}.&&
\end{array}
\end{eqnarray}
The eigenvalues of the central charges are denoted by
$H,C,C^{\dag}$. The charge $H$ is Hermitian and the charges
$C,C^{\dagger}$ and the generators $\mathbb{Q},\mathbb{G}$ are
conjugate to each other.\smallskip

For computational purposes, it proves worthwhile to consider
representations of the algebra in the superspace formalism.
Consider the vector space of analytic functions of two bosonic
variables $w_{1,2}$ and two fermionic variables $\theta_{3,4}$.
Since we are dealing with analytic functions we can expand any
such function $\Phi(w,\theta)$:
\begin{eqnarray}
\Phi(w,\theta) &=&\sum_{\ell=0}^{\infty}\Phi_{\ell}(w,\theta),\nonumber\\
\Phi_{\ell} &=& \phi^{a_{1}\ldots a_{\ell}}w_{a_{1}}\ldots
w_{a_{\ell}} +\phi^{a_{1}\ldots a_{\ell-1}\alpha}w_{a_{1}}\ldots
w_{a_{\ell-1}}\theta_{\alpha}+\nonumber\\
&&\phi^{a_{1}\ldots a_{\ell-2}\alpha\beta}w_{a_{1}}\ldots
w_{a_{\ell-2}}\theta_{\alpha}\theta_{\beta}.
\end{eqnarray}
The representation that describes $\ell$-particle bound states of
the light-cone string theory on $\ads$ of centrally extended
$\mathfrak{su}(2|2)$ is $4\ell$ dimensional. It is realized on a
graded vector space with basis $|e_{a_{1}\ldots a_{\ell}}\rangle,
|e_{a_{1}\ldots a_{\ell-1}\alpha}\rangle,|e_{a_{1}\ldots
a_{\ell-2}\alpha\beta}\rangle$, where $a_{i}$ are bosonic indices
and $\alpha,\beta$ are fermionic indices and each of the basis
vectors is totally symmetric in the bosonic indices and
anti-symmetric in the fermionic indices
\cite{Arutyunov:2008,Beisert:2006qh,Dorey:2006dq}. In terms of the
above analytic functions, the basis vectors of the totally
symmetric representation can evidently be identified
$|e_{a_{1}\ldots a_{\ell}}\rangle \leftrightarrow w_{a_{1}}\ldots
w_{a_{\ell}},|e_{a_{1}\ldots a_{\ell-1}\alpha}\rangle
\leftrightarrow w_{a_{1}}\ldots w_{a_{\ell-1}}\theta_{\alpha}$ and
$|e_{a_{1}\ldots a_{\ell-1}\alpha\beta}\rangle \leftrightarrow
w_{a_{1}}\ldots w_{a_{\ell-2}}\theta_{\alpha}\theta_{\beta}$
respectively. In other words, we find the atypical totally
symmetric representation describing $\ell$-particle bound states
when we restrict to terms $\Phi_{\ell}$. For later convenience, we
introduce the notation $W^{(m)}_{1^{i}2^{j}3^{k}4^{l}} \equiv
(w^{(m)}_{1})^{i}(w^{(m)}_{2})^{j}(\theta^{(m)}_{3})^{k}(\theta_{4}^{(m)})^{l}$,
where $(m)$ denotes different copies of the representation.
Clearly, for an $\ell$-particle bound state representation, this
means that $i+j+k+l=\ell$ and $k,l = 0,1$.
\smallskip

In this representation the algebra generators can be written in
differential operator form in the following way
\begin{eqnarray}\label{eqn;AlgDiff}
\begin{array}{lll}
  \mathbb{L}_{a}^{\ b} = w_{a}\frac{\partial}{\partial w_{b}}-\frac{1}{2}\delta_{a}^{b}w_{c}\frac{\partial}{\partial w_{c}}, &\qquad& \mathbb{R}_{\alpha}^{\ \beta} = \theta_{\alpha}\frac{\partial}{\partial \theta_{\beta}}-\frac{1}{2}\delta_{\alpha}^{\beta}\theta_{\gamma}\frac{\partial}{\partial \theta_{\gamma}}, \\
  \mathbb{Q}_{\alpha}^{\ a} = a \theta_{\alpha}\frac{\partial}{\partial w_{a}}+b\epsilon^{ab}\epsilon_{\alpha\beta} w_{b}\frac{\partial}{\partial \theta_{\beta}}, &\qquad& \mathbb{G}_{a}^{\ \alpha} = d w_{a}\frac{\partial}{\partial \theta_{\alpha}}+c\epsilon_{ab}\epsilon^{\alpha\beta} \theta_{\beta}\frac{\partial}{\partial w_{b}}
\end{array}
\end{eqnarray}
and the central charges are
\begin{eqnarray}
\begin{array}{ll}
 \mathbb{C} = ab \left(w_{a}\frac{\partial}{\partial w_{a}}+\theta_{\alpha}\frac{\partial}{\partial
 \theta_{\alpha}}\right)& \mathbb{C}^{\dag} = cd \left(w_{a}\frac{\partial}{\partial w_{a}}+\theta_{\alpha}\frac{\partial}{\partial
 \theta_{\alpha}}\right)\\
 \mathbb{H}= (ad +bc)\left(w_{a}\frac{\partial}{\partial w_{a}}+\theta_{\alpha}\frac{\partial}{\partial
 \theta_{\alpha}}\right).
\end{array}
\end{eqnarray}
To form a representation, the parameters $a,b,c,d$ satisfy the
condition $ad-bc=1$. The central charges become $\ell$ dependent:
\begin{eqnarray}
H= \ell(ad+bc),\qquad C =\ell ab , \qquad C^{\dag} =\ell cd.
\end{eqnarray}
The parameters $a,b,c,d$ can be expressed in terms of the particle
momentum $p$ and the coupling $g$:
\begin{eqnarray}
\begin{array}{lll}
  a = \sqrt{\frac{g}{2\ell}}\eta & \quad & b = \sqrt{\frac{g}{2\ell}}
\frac{i\zeta}{\eta}\left(\frac{x^{+}}{x^{-}}-1\right) \\
  c = -\sqrt{\frac{g}{2\ell}}\frac{\eta}{\zeta x^{+}} & \quad &
d=\sqrt{\frac{g}{2\ell}}\frac{x^{+}}{i\eta}\left(1-\frac{x^{-}}{x^{+}}\right),
\end{array}
\end{eqnarray}
where the parameters $x^{\pm}$ satisfy
\begin{eqnarray}
x^{+} +
\frac{1}{x^{+}}-x^{-}-\frac{1}{x^{-}}=\frac{2i\ell}{g},\qquad
\frac{x^{+}}{x^{-}} = e^{ip}
\end{eqnarray}
and the parameters $\eta$ are given by
\begin{eqnarray}\label{eqn;ScatteringBasis}
\eta = e^{i\xi}\eta(p),\qquad \eta(p)=
e^{\frac{i}{4}p}\sqrt{ix^{-}-ix^{+}},\qquad \zeta = e^{2i\xi}.
\end{eqnarray}
The fundamental representation, which is used in the derivation of
the S-matrix scattering fundamental multiplets
\cite{Beisert:2005tm,Arutyunov:2006yd} is obtained by taking
$\ell=1$.

\subsection{Yangians, coproducts and the S-matrix}

The double Yangian $DY(\mathfrak{g})$ of a (simple) Lie algebra
$\mathfrak{g}$ is a deformation of the universal enveloping
algebra $U(\mathfrak{g}[u,u^{-1}])$ of the loop algebra
$\mathfrak{g}[u,u^{-1}]$. The Yangian is generated by level $n$
generators $\mathbb{J}^{A}_{n},\ n\in\mathbb{Z}$ that satisfy the
commutation relations
\begin{eqnarray}
\ [\mathbb{J}^{A}_{m},\mathbb{J}^{B}_{n}  ]  = F^{AB}_{C}
\mathbb{J}^{C}_{m+n} + \mathcal{O}(\hbar),
\end{eqnarray}
where $F^{AB}_{C}$ are the structure constants of $\mathfrak{g}$.
The level-0 generators $\mathbb{J}_{0}^{A}$ span the Lie-algebra.
The $\su(2|2)$ Yangian can be supplied with a coproduct in the
following way \cite{Gomez:2006va,Plefka:2006ze,Beisert:2007ds}
\begin{eqnarray}\label{eqn;CoProdYang}
\Delta (\mathbb{J}^{A}_{n})  &=& \mathbb{J}^{A}_{n}\otimes 1 +
\mathcal{U}^{[A]}\otimes \mathbb{J}^{A}_{n} + \frac{\hbar }{2}
\sum_{m=0}^{n-1} F_{BC}^{A}
\mathbb{J}^{B}_{n-1-m}\mathcal{U}^{[C]}\otimes \mathbb{J}^{C}_{m} +\mathcal{O}(\hbar^{2})\nonumber\\
\Delta(\mathcal{U})&=&\mathcal{U}\otimes \mathcal{U},
\end{eqnarray}
where $\mathcal{U}$ comprises a so-called braiding
factor.\smallskip

An important representation of a Yangian is the evaluation
representation. This representation consists of states
$|u\rangle$, with action $\mathbb{J}^{A}_{n}|u\rangle =
u^{n}\mathbb{J}^{A}_{0}|u\rangle$. In this representation the
coproduct structure is fixed in terms of the coproducts of
$\mathbb{J}_0,\mathbb{J}_1$. For the remainder of this paper we
will work in this representation and identify
$\mathbb{J}_1\equiv\hat{\mathbb{J}} = \frac{g}{2i}u\mathbb{J}$ for
the $\su(2|2)$ Yangian. One also finds that $u$ is actually
dependent on the parameters $x^{\pm}$ via $u_{j} =
x_{j}^{+}+\frac{1}{x_{j}^{+}} - \frac{i\ell_{j}}{g}$.\smallskip

The S-matrix is a map between the following representations:
\begin{eqnarray}
\S:~~ \V_{\ell_{1}}(p_{1},e^{ip_{2}})\otimes \V_{\ell_{2}}(p_{2},1
)\longrightarrow \V_{\ell_{1}}(p_{1},1 )\otimes
\V_{\ell_{2}}(p_{2},e^{ip_{1}}),
\end{eqnarray}
where $\V_{\ell_{i}}(p_{i},e^{2i\xi})$ is the $\ell_{i}$-bound
state representation with parameters $a_{i},b_{i},c_{i},d_{i}$
with the explicit choice of $\zeta = e^{2i\xi}$. This specific
choice removes the braid factor $\mathcal{U}$ from the discussion
\cite{Arutyunov:2006yd}.\smallskip

The bound state S-matrices are now fixed, up to a overall phase,
by requiring invariance under the coproducts of the (Yangian)
symmetry generators
\begin{eqnarray}
\S~\Delta(\mathbb{J}^{A})&=&\Delta^{op}(\mathbb{J}^{A})~\S\nonumber\\
\S~\Delta(\hat{\mathbb{J}}^{A})&=&\Delta^{op}(\hat{\mathbb{J}}^{A})~\S,
\end{eqnarray}
where $\Delta^{op} = P \Delta$ with $P$ the graded
permutation.\smallskip

For completeness and future reference, we will give the explicit
formulas of the different coproducts and of the parameters
$a_i,b_i,c_i,d_i$. First, the $\su(2|2)$ operators:
\begin{eqnarray}\label{eqn;CoprodSymm}
\Delta(\mathbb{J}_{0}^{A}) =\mathbb{J}_{1;0}^{A} +
\mathbb{J}_{2;0}^{A}.
\end{eqnarray}
The coproducts of the Yangian generators are
\begin{eqnarray}\label{eqn;CoProdYangDiff}
\Delta(\hat{\mathbb{L}}^{a}_{\ b}) &=& \hat{\mathbb{L}}^{\
a}_{1;b} + \hat{\mathbb{L}}^{\ a}_{2;b} + \frac{1}{2}\mathbb{L}^{\
c}_{1;b}\mathbb{L}^{\ a}_{2;c}-\frac{1}{2} \mathbb{L}^{\
a}_{1;c}\mathbb{L}^{\ c}_{2;b}-\frac{1}{2} \mathbb{G}^{\
\gamma}_{1;b}\mathbb{Q}^{\ a}_{2;\gamma}-\frac{1}{2} \mathbb{Q}^{\
a}_{1;\gamma}\mathbb{G}^{\
\gamma}_{2;b}\nonumber\\
&&+\frac{1}{4}\delta^{a}_{b}\mathbb{G}^{\
\gamma}_{1;c}\mathbb{Q}^{\ c}_{2;\gamma}+\frac{1}{4}\delta^{a}_{b}
\mathbb{Q}^{\ c}_{1;\gamma}\mathbb{G}^{\
\gamma}_{2;c}\nonumber\\
\Delta(\hat{\mathbb{R}}^{\alpha}_{\ \beta}) &=&
\hat{\mathbb{R}}^{\ \alpha}_{1;\beta} + \hat{\mathbb{R}}^{\
\alpha}_{2;\beta} - \frac{1}{2}\mathbb{R}^{\
\gamma}_{1;\beta}\mathbb{R}^{\ \alpha}_{2;\gamma}+\frac{1}{2}
\mathbb{R}^{\ \alpha}_{1;\gamma}\mathbb{R}^{\
\gamma}_{2;\beta}+\frac{1}{2} \mathbb{G}^{\
\alpha}_{1;c}\mathbb{Q}^{\ c}_{2;\beta}+\frac{1}{2} \mathbb{Q}^{\
c}_{1;\beta}\mathbb{G}^{\
\alpha}_{2;c}\nonumber\\
&&-\frac{1}{4}\delta^{\alpha}_{\beta}\mathbb{G}^{\
\gamma}_{1;c}\mathbb{Q}^{\
c}_{2;\gamma}-\frac{1}{4}\delta^{\alpha}_{\beta} \mathbb{Q}^{\
c}_{1;\gamma}\mathbb{G}^{\
\gamma}_{2;c}\\
\Delta(\hat{\mathbb{Q}}^{a}_{\ \beta}) &=& \hat{\mathbb{Q}}^{\
a}_{1;\beta} + \hat{\mathbb{Q}}^{\ a}_{2;\beta} -
\frac{1}{2}\mathbb{R}^{\ \gamma}_{1;\beta}\mathbb{Q}^{\
a}_{2;\gamma}+\frac{1}{2} \mathbb{Q}^{\ a}_{1;\gamma}\mathbb{R}^{\
\gamma}_{2;\beta} -\frac{1}{2} \mathbb{L}^{\ a}_{1;c}\mathbb{Q}^{\
c}_{2;\beta}+\frac{1}{2} \mathbb{Q}^{\
c}_{1;\beta}\mathbb{L}^{\ a}_{2;c}\nonumber\\
&&-\frac{1}{4}\mathbb{H}_{1}\mathbb{Q}^{\
a}_{2;\beta}+\frac{1}{4}\mathbb{Q}^{\ a}_{1;\beta}\mathbb{H}_{2} +
\frac{1}{2}\epsilon_{\beta\gamma}\epsilon^{ad}\mathbb{C}_{1}\mathbb{G}^{\
\gamma}_{2;d}-\frac{1}{2}\epsilon_{\beta\gamma}\epsilon^{ad}\mathbb{G}^{\
\gamma}_{1;d}\mathbb{C}_{2}\nonumber\\
\Delta(\hat{\mathbb{G}}^{\alpha}_{\ b}) &=& \hat{\mathbb{G}}^{\
\alpha}_{1;b} + \hat{\mathbb{G}}^{\ \alpha}_{2;b} +
\frac{1}{2}\mathbb{L}^{\ c}_{1;b}\mathbb{G}^{\ \alpha}_{2;c}-
\frac{1}{2}\mathbb{G}^{\ \alpha}_{1;c}\mathbb{L}^{\ c}_{2;b}
+\frac{1}{2} \mathbb{R}^{\ \alpha}_{1;\gamma}\mathbb{G}^{\
\gamma}_{2;b} -\frac{1}{2} \mathbb{G}^{\
\gamma}_{1;b}\mathbb{R}^{\ \alpha}_{2;\gamma}\nonumber\\
&&+\frac{1}{4}\mathbb{H}_{1}\mathbb{G}^{\ \alpha}_{2;b}-
\frac{1}{4}\mathbb{G}^{\ \alpha}_{1;b}\mathbb{H}_{2} -
\frac{1}{2}\epsilon_{bc}\epsilon^{\alpha\gamma}\mathbb{C}^{\dag}_{1}\mathbb{Q}^{\
c}_{2;\gamma}
+\frac{1}{2}\epsilon_{bc}\epsilon^{\alpha\gamma}\mathbb{Q}^{\
c}_{1;\gamma}\mathbb{C}^{\dag}_{2}\nonumber
\end{eqnarray}
and for the central charges
\begin{eqnarray}
\Delta(\hat{\mathbb{H}}) &=& \hat{\mathbb{H}}_{1} +\hat{\mathbb{H}}_{2}+ \mathbb{C}_{1}\mathbb{C}^{\dag}_{2}-\mathbb{C}^{\dag}_{1}\mathbb{C}_{2}\nonumber\\
\Delta(\hat{\mathbb{C}}) &=& \hat{\mathbb{C}}_{1}+\hat{\mathbb{C}}_{2}+\frac{1}{2} \mathbb{H}_{1}\mathbb{C}_{2}-\frac{1}{2} \mathbb{C}_{1}\mathbb{H}_{2}\\
\Delta(\hat{\mathbb{C}}^{\dag}) &=&
\hat{\mathbb{C}}^{\dag}_{1}+\hat{\mathbb{C}}^{\dag}_{2}+\frac{1}{2}
\mathbb{H}_{1}\mathbb{C}^{\dag}_{2}-\frac{1}{2}
\mathbb{C}^{\dag}_{1}\mathbb{H}_{2}\nonumber.
\end{eqnarray}
The product is ordered, e.g. $\mathbb{Q}_{1}\mathbb{Q}_{2}$ means
first apply $\mathbb{Q}_{2}$, then $\mathbb{Q}_{1}$ (as
differential operators). Finally, the coefficients used in
$\Delta$ are given by:
\begin{eqnarray}
\begin{array}{lll}
  a_{1} = \sqrt{\frac{g}{2\ell_{1}}}\eta_{1} & ~ & b_{1} =
 -i e^{i p_{2}}\sqrt{\frac{g}{2\ell_{1}}}~
\frac{1}{\eta_{1}}\left(\frac{x_{1}^{+}}{x_{1}^{-}}-1\right) \\
  c_{1} = -e^{-i p_{2}}\sqrt{\frac{g}{2\ell_{1}}}\frac{\eta_{1}}{ x_{1}^{+}} & ~ &
d_{1}=i\sqrt{\frac{g}{2\ell_{1}}}\frac{x_{1}^{+}}{\eta_{1}}\left(\frac{x_{1}^{-}}{x_{1}^{+}}-1\right)\\
\eta_{1} =
e^{i\frac{p_{1}}{4}}e^{i\frac{p_{2}}{2}}\sqrt{ix^{-}_{1}-ix^{+}_{1}}&~&~\\
~ &~& ~ \\
  a_{2} = \sqrt{\frac{g}{2\ell_{2}}}\eta_{2} & ~ & b_{2} = -i\sqrt{\frac{g}{2\ell_{2}}}
\frac{1}{\eta_{2}}\left(\frac{x_{2}^{+}}{x_{2}^{-}}-1\right) \\
  c_{2} = -\sqrt{\frac{g}{2\ell_{2}}}\frac{\eta_{2}}{x_{2}^{+}} & ~ &
d_{2}=i\sqrt{\frac{g}{2\ell_{2}}}\frac{x_{2}^{+}}{i\eta_{2}}\left(\frac{x_{2}^{-}}{x_{2}^{+}}-1\right)\\
\eta_{2} = e^{i\frac{p_{2}}{4}}\sqrt{ix^{-}_{2}-ix^{+}_{2}}&~&~
\end{array}
\end{eqnarray}
The coefficients in $\Delta^{op}$ are given by:
\begin{eqnarray}
\begin{array}{lll}
  a^{op}_{1} = \sqrt{\frac{g}{2\ell_{1}}}\eta^{op}_{1} & ~ & b^{op}_{1} = -i\sqrt{\frac{g}{2\ell_{1}}}
\frac{1}{\eta^{op}_{1}}\left(\frac{x_{1}^{+}}{x_{1}^{-}}-1\right) \\
  c^{op}_{1} = -\sqrt{\frac{g}{2\ell_{1}}}\frac{\eta^{op}_{1}}{x_{1}^{+}} & ~ &
d^{op}_{1}=i\sqrt{\frac{g}{2\ell_{1}}}\frac{x_{1}^{+}}{i\eta^{op}_{1}}\left(\frac{x_{1}^{-}}{x_{1}^{+}}-1\right)\\
\eta^{op}_{1} =
e^{i\frac{p_{1}}{4}}\sqrt{ix^{-}_{1}-ix^{+}_{1}}&~&~\nonumber\\
 ~ &~& ~ \\
  a^{op}_{2} = \sqrt{\frac{g}{2\ell_{2}}}\eta^{op}_{2} & ~ &
b^{op}_{2} =
 -i e^{i p_{1}}\sqrt{\frac{g}{2\ell_{2}}}~
\frac{1}{\eta^{op}_{2}}\left(\frac{x_{2}^{+}}{x_{2}^{-}}-1\right) \\
  c^{op}_{2} = -e^{-i p_{1}}\sqrt{\frac{g}{2\ell_{2}}}\frac{\eta^{op}_{2}}{ x_{2}^{+}} & ~ &
d^{op}_{2}=i\sqrt{\frac{g}{2\ell_{2}}}\frac{x_{2}^{+}}{\eta^{op}_{2}}\left(\frac{x_{2}^{-}}{x_{2}^{+}}-1\right)\\
\eta^{op}_{2} =
e^{i\frac{p_{2}}{4}}e^{i\frac{p_{1}}{2}}\sqrt{ix^{-}_{2}-ix^{+}_{2}}&~&~
\end{array}
\end{eqnarray}
The non-trivial braiding factors are all hidden in the parameters
of the four representations involved.\smallskip

\section{The $\su(2|2)$ coordinate Bethe Ansatz}

In this section we will briefly discuss the coordinate Bethe
Ansatz for the $\su(2|2)$ string S-matrix as done in
\cite{Beisert:2005tm,Leeuw:2007uf}.

\subsection{Formalism}

Consider $K^{\mathrm{I}}$ excitations with momentum $p_{1},\ldots,
p_{K^{\mathrm{I}}}$. We divide our space into regions
$\mathcal{P}|\mathcal{Q}$, with $\mathcal{P},\mathcal{Q}$
permutations. In the sector where the particle coordinates are
ordered as
$x_{\mathcal{Q}_1}<\ldots<x_{\mathcal{Q}_{K^{\mathrm{I}}}}$ the
ansatz for the wave function is given by
\begin{eqnarray}
|p_{1},\ldots,p_{K^{\mathrm{I}}}\rangle = \sum_{\mathcal{P}}\int
dx \left\{A^{\mathcal{P}|\mathcal{Q}}_{a_{1}\ldots
a_{K^{\mathrm{I}}}}e^{i
p_{\mathcal{P}_i}x_{\mathcal{Q}_i}}\right\}\phi^{a_1}(x_1)\ldots
\phi^{a_K}(x_{K^{\mathrm{I}}})
\end{eqnarray}
The indices $a_{i}$ denote the type of the $i$th particle and
$\phi^{a_i}(x_i)$ is a creation operator that creates a particle
of type $a_i$ at position $x_i$. In general when the system
contains bound states, the indices $a_i$ also run over bound
states. The different sectors are related to each other by
permutations and S-matrices. To be more precise
\begin{eqnarray}\label{eqn;compat}
A^{\mathcal{P}|\mathcal{Q}} &=& \S_{i,j}
A^{\mathcal{P}'|\mathcal{Q}'},
\end{eqnarray}
where $P'|Q'$ are the permutations obtained from $P|Q$ by
interchanging the neighboring $i$th and $j$th particles. To make
this transformation property more explicit we will write
\begin{eqnarray}
A^{\mathcal{P}|\mathcal{Q}}_{a_{1}\ldots a_{K}} \equiv
A^{\mathcal{P}|\mathcal{Q}}_{a_{1}\ldots a_{K}}w_{a_1}\ldots
w_{a_k},
\end{eqnarray}
where we do not sum over repeated indices. Periodicity of the
total wave function is then formulated in the following Bethe
equations
\begin{eqnarray}\label{eqn;BAE}
\S_{k}A^{\mathcal{P}|\mathcal{Q}}:=\S_{k k-1}\ldots \S_{k
K^{\rm{I}}}\S_{k 1}\ldots \S_{kk+1}A^{\mathcal{P}|\mathcal{Q}} =
e^{ip_{k}L}A^{\mathcal{P}|\mathcal{Q}}.
\end{eqnarray}
The Bethe ansatz now consists of making a ansatz for the
coefficients $A^{\mathcal{P}|\mathcal{Q}}$ in such a way that they
solve (\ref{eqn;compat}). This ansatz is then plugged in the above
formula and the explicit Bethe equations can be read
off.\smallskip

Obviously, since any scattering process reduces to a product of
two body scattering processes, we only need to consider two sites
of the wave function. In what follows we will solve the for the
coefficients $A^{\mathcal{P}|\mathcal{Q}}_{i_{1}\ldots
i_{K^{\mathrm{I}}}}$.

\subsection{Solving the coefficients}

We will now solve the coefficients for the $\su(2|2)$ S-matrix for
fundamental representations. In order make the derivation more
transparent, we explicitly identify the coefficients $A^{P|Q}$
with states $|\cdot\rangle\in
\V_{1}(p_1)\otimes\ldots\otimes\V_{1}(p_{K^{\mathrm{I}}})$. From
now on we will omit the explicit mentioning of the sector
$\mathcal{P}|\mathcal{Q}$ we are in. The different sectors are
related via relabelling of momenta and interchanging of particle
positions and hence the coefficients should be well-defined in the
sense that they should respect this. From (\ref{eqn;compat}), we
see that in order to be well-defined, we must have that
\begin{eqnarray}\label{eqn;WellDefined}
\mathbb{S}|\phi\rangle = |\phi\rangle_{\pi},
\end{eqnarray}
where $|\phi\rangle_{\pi}$ is the coefficient as constructed in
the sector $\mathcal{P}^{\prime}|\mathcal{Q}^{\prime}$. \smallskip

Let us first define the vacuum:
\begin{eqnarray}
|0\rangle = W_{1}^{(1)}\ldots W_{1}^{(K)}.
\end{eqnarray}
One can also take other vacua but this leads to equivalent
equations. If we, for the moment, consider the undressed S-matrix,
we find that
\begin{eqnarray}
\S|0\rangle = |0\rangle.
\end{eqnarray}
In the light of (\ref{eqn;WellDefined}), we indeed find that this
vacuum corresponds to a vacuum in all sectors
$\mathcal{P}|\mathcal{Q}$ and is well defined.\smallskip

The complete dressing phase $S_0(p_i,p_j)$ can be easily
incorporated by a rescaling
\begin{eqnarray}
|0\rangle = \frac{W_{1}^{(1)}\ldots
W_{1}^{(K)}}{\sqrt{\prod_{i\neq j}S_0(p_i,p_j)}}.
\end{eqnarray}
This works because $S_0(p_j,p_i) = S_0(p_j,p_i)^{-1}$. To avoid
making the formulas too cumbersome, we will restrict to the
undressed S-matrix in the remainder of the paper. We will, of
course, include the dressing phase in the full Bethe
equations.
\smallskip

\begin{figure}
  \centering
\includegraphics{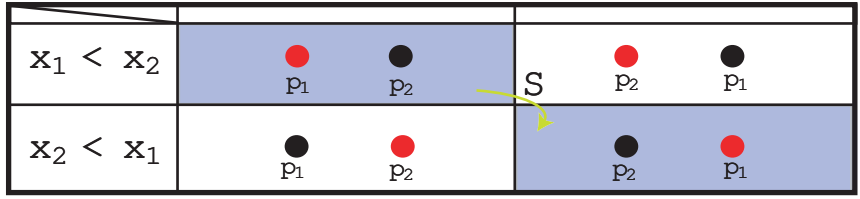}
  \caption{Example of the construction of the coefficients in the different sectors. The black dots
  correspond to the bosons of the vacuum, $W_1$, and the red dots
  correspond to the inserted fermions. The construction is
  well-defined if, by acting on the coefficient in the upper left
  quadrant by the S-matrix, one obtains the coefficient in the
  lower right quadrant.
  }\label{Fig;WellDefined}
\end{figure}

The next thing to consider is the case where we have a fermion in
this vacuum (the case in which the other boson is inserted is
treated later on). We make an ansatz of the following form:
\begin{eqnarray}
|\alpha\rangle:= \sum_{i} \Psi_{i}(y) W^{(1)}_{1}\ldots
W^{(i)}_{\alpha}\ldots W_{1}^{(K)},\qquad \Psi_{k}(y) =
f(y,p_{k})\prod_{l<k} S^{\mathrm{II,I}}(y,p_{l}).
\end{eqnarray}
We will denote $S(y,p)\equiv S^{\mathrm{II,I}}(y,p)$. We must
check whether this construction is well defined in the sense that
it respects (\ref{eqn;WellDefined}). How this works is
schematically depicted in Figure \ref{Fig;WellDefined}.\smallskip

By the factorization property of the S-matrix, it suffices to
restrict to a two particle state and a two particle S-matrix. For
this case (\ref{eqn;WellDefined}) gives the following equations
\begin{eqnarray}\label{eqn;level1}
\frac{e^{i\frac{p_1}{2}}}{e^{i\frac{p_2}{2}}}\frac{\eta(p_2)}{\eta(p_{1})}\frac{x^{+}_{1}-x^{-}_{1}}{x^{+}_{1}-x^{-}_{2}}~f(y,p_1)
+
 e^{i\frac{p_1}{2}}\frac{x^{-}_{1}-x^{-}_{2}}{x^{+}_{1}-x^{-}_{2}}~ f(y,p_2)S(y,p_1) &=& f(y,p_2)\nonumber\\
e^{-i\frac{p_2}{2}}\frac{x^{+}_{1}-x^{+}_{2}}{x^{+}_{1}-x^{-}_{2}}~f(y,p_1)
+
\frac{\eta(p_{1})}{\eta(p_{2})}\frac{x^{+}_{2}-x^{-}_{2}}{x^{+}_{1}-x^{-}_{2}}~
f(y,p_2)S(y,p_1) &=& f(y,p_1) S(y,p_2).
\end{eqnarray}
These equations can be solved explicitly and the solution is given
by:
\begin{eqnarray}\label{eqn;SolutionfS}
f(y,p_{k}) &=& \eta(p_{k}) \sqrt{\frac{x^-_k}{x^+_k}}\frac{y}{y- x_{k}^{-}}\sqrt{\frac{g \ell_{k}}{2}}\nonumber\\
S(y,p_{k}) &=& \sqrt{\frac{x^-_k}{x^+_k}}\frac{y-x^{+}_{k}}{y-
x_{k}^{-}},
\end{eqnarray}
where $y$ enters as a integration constant. With a modest amount
of foresight, we choose the overall normalization of $f$ to be
dependent on the bound state number $\ell_{k}$.
\smallskip

Actually this solution is the unique solution of
(\ref{eqn;WellDefined}) for one fermion in the vacuum up to
overall normalization (and hence the only well-defined coefficient
for this case). This can easily be seen from (\ref{eqn;level1}).
The general wave function is of the form
\begin{eqnarray}
|\alpha\rangle:= A_1 W^{(1)}_{\alpha}\ldots W_{1}^{(K)} + A_2
W^{(1)}_{1}W^{(2)}_{\alpha}\ldots W_{1}^{(K)}+\ldots .
\end{eqnarray}
Since we are only interested in solving (\ref{eqn;BAE}), the
normalization of the state is irrelevant. Let us therefore pick a
convenient normalization, $A_1 \equiv f(p_{1})$, then by the first
equation of (\ref{eqn;level1}) one finds $A_{2}$ in terms of
$A_{1}$. This indeed fixes $A_{2}$ as $f(p_2)S(p_1)$ and by
induction the rest of the coefficients also follow. Finally, let
us stress that by fixing the normalization, one automatically
finds from the S-matrix that the other coefficients of the wave
function are of a factorized form.
\smallskip

The problem becomes more involved upon inserting two excitations.
In this case, it appears that the ansatz for the coefficient can
be written as
\begin{eqnarray}
|\alpha\beta\rangle = |\alpha\beta\rangle_{y_1 y_2} +
\S^{\mathrm{II}}|\alpha\beta\rangle_{y_1 y_2},
\end{eqnarray}
where we introduce a new S-matrix $\S^{\mathrm{II}}$
\begin{eqnarray}
\S^{\mathrm{II}}|\alpha\beta\rangle_{y_1 y_2} &=&
M(y_1,y_2)|\alpha\beta\rangle_{y_2 y_1} +
N(y_1,y_2)|\beta\alpha\rangle_{y_2 y_1}.
\end{eqnarray}
Explicitly, it is given by
\begin{eqnarray}
|\alpha\beta\rangle &=&
\sum_{k<l}\Psi_k(y_1)\Psi_l(y_2)W^{(1)}_{1}\ldots
W^{(k)}_{\alpha}\ldots W^{(l)}_{\beta}\ldots
W_{1}^{(K)}+\nonumber\\
&&+\S^{\mathrm{II}}\sum_{k<l}\Psi_k(y_1)\Psi_l(y_2)W^{(1)}_{1}\ldots
W^{(k)}_{\alpha}\ldots W^{(l)}_{\beta}\ldots W_{1}^{(K)}+\\
&&+\epsilon^{\alpha\beta}\sum_{k}\Psi_k(y_1)\Psi_k(y_2)h(y_1,y_2,p_k)W^{(1)}_{1}\ldots
W^{(k)}_{2}\ldots W_{1}^{(K)}\nonumber
\end{eqnarray}
When we restrict to just two sites, the wave function splits into
the sum of a wave function with just one fermion. By construction,
this piece is exactly as described above. The other piece contains
two excitations and is given by
\begin{eqnarray}
|\alpha\beta\rangle &=& \left\{f(y_1,p_1) f(y_2,p_2)S(y_2,p_1) + M f(y_2,p_1) f(y_1,p_2)S(y_1,p_1)\right\} W^{(1)}_{\alpha}W^{(2)}_{\beta}\nonumber\\
&& + N f(y_2,p_1) f(y_1,p_2)S(y_1,p_1)W^{(1)}_{\beta}W^{(2)}_{\alpha}\nonumber \\
&& + \epsilon^{\alpha\beta}h(y_1,y_2,p_1)f(y_2,p_1) f(y_1,p_1)W^{(1)}_{2}W^{(2)}_{1}\\
&& + \epsilon^{\alpha\beta}h(y_1,y_2,p_2)f(y_2,p_2)
f(y_1,p_2)S(y_2,p_1) S(y_1,p_1)W^{(1)}_{1}W^{(2)}_{2}\nonumber
\end{eqnarray}
Plugging this into (\ref{eqn;WellDefined}) again allows one to
find the explicit (unique) solutions of the unknown functions:
\begin{eqnarray}\label{eqn;TwoExcSolution}
M(y_1,y_2) &=& \frac{2 i/g  }{y_1+\frac{1}{y_1}-y_2-\frac{1}{y_2}-\frac{2 i}{g}}\nonumber\\
N(y_1,y_2) &=&
-\frac{y_1+\frac{1}{y_1}-y_2-\frac{1}{y_2}}{y_1+\frac{1}{y_1}-y_2-\frac{1}{y_2}-\frac{2 i}{g}}\\
h(y_1,y_2,p_k) &=& \frac{i}{\ell_{k}\eta(p_{k})^2}~ \frac{y_1 y_2
- x_{k}^{+}x_{k}^{-}}{y_1
y_2}~\frac{x_{k}^{+}-x_{k}^{-}}{x^{-}_{k}}~
\frac{y_1-y_2}{y_1+\frac{1}{y_1}-y_2-\frac{1}{y_2}-\frac{2
i}{g}}\nonumber
\end{eqnarray}
In this process, we introduced a new S-matrix and we also insist
on naturalness of the wave function with respect to this S-matrix,
$\S^{\mathrm{II}}$. One can repeat the above procedure to deal
with this. For details we refer to
\cite{Leeuw:2007uf,Beisert:2005tm}. In this process we are again
lead to introduce new functions $S^{\mathrm{II,II}},f^{(2)},
S^{\mathrm{III},\mathrm{II}}, S^{\mathrm{III,III}}$. The result of
this consideration is
\begin{eqnarray}
S^{\mathrm{II,II}}&=&-M-N=1\nonumber\\
f^{(2)}(w,y_{k})&=&\frac{w-\frac{i}{g}}{w-v_{k}-\frac{i}{g}}\nonumber\\
S^{\mathrm{III},\mathrm{II}}(w,y_{k})&=&\frac{w-v_{k}+\frac{i}{g}}{w-v_{k}-\frac{i}{g}}\\
S^{\mathrm{III,III}}(w_{1},w_{2}) &=&
\frac{w_{1}-w_{2}-\frac{2i}{g}}{w_{1}-w_{2}+\frac{2i}{g}}\nonumber,
\end{eqnarray}
The Bethe equations (\ref{eqn;BAE}) are formulated in terms of the
factors used in the ansatz in the following way
\begin{eqnarray}\label{RealBAE} e^{iL_{A,k}}=
{\prod_{B=\mathrm{I}}^{\mathrm{III}}\prod_{l=1}^{K^{B}}}
S^{BA}(x^{B}_{l},x^{A}_{k}),\quad (B,l)\neq (A,k)
\end{eqnarray}
where $A,B$ denote the different levels and $S^{AB}$ can be seen
as the S-matrix describing scattering at different levels.
Moreover, $e^{iL_{\mathrm{I},k}}=e^{-iLp_{k}}$ and
$e^{iL_{\mathrm{II},k}}=e^{iL_{\mathrm{III},k}}=1$ are
phases.\smallskip

By putting all of this together one obtains the well-known Bethe
equations describing the asymptotic spectrum of the $\ads$
superstring:
\begin{eqnarray}
e^{ip_{k}L}&=& \prod_{l=1,l\neq k
}^{K^{\mathrm{I}}}\left[S_{0}(p_{k},p_{l})\frac{x_k^+-x_l^-}{x_k^--x_l^+}\sqrt{\frac{x_l^+x_k^-}{x_l^-x_k^+}}\right]^2\prod_{\alpha=1}^{2}\prod_{l=1}^{K_{(\alpha)}^{\mathrm{II}}}\frac{{x_{k}^{-}-y^{(\alpha)}_{l}}}{x_{k}^{+}-y^{(\alpha)}_{l}}\sqrt{\frac{x^+_k}{x^-_k}}\nonumber \\
1&=&\prod_{l=1}^{K^{\mathrm{I}}}\frac{y^{(\alpha)}_{k}-x^{+}_{l}}{y^{(\alpha)}_{k}-x^{-}_{l}}\sqrt{\frac{x^-_k}{x^+_k}}
\prod_{l=1}^{K_{(\alpha)}^{\mathrm{III}}}\frac{y_{k}^{(\alpha)}+\frac{1}{y_{k}^{(\alpha)}}-w_{l}^{(\alpha)}+\frac{i}{g}}{y_{k}^{(\alpha)}+\frac{1}{y_{k}^{(\alpha)}}-w_{l}^{(\alpha)}-\frac{i}{g}}\\
1&=&\prod_{l=1}^{K_{(\alpha)}^{\mathrm{II}}}\frac{w_{k}^{(\alpha)}-y_{k}^{(\alpha)}-\frac{1}{y_{k}^{(\alpha)}}+\frac{i}{g}}{w_{k}^{(\alpha)}-y_{k}^{(\alpha)}-\frac{1}{y_{k}^{(\alpha)}}-\frac{i}{g}}
\prod_{l\neq
k}^{K_{(\alpha)}^{\mathrm{III}}}\frac{w_{k}^{(\alpha)}-w_{l}^{(\alpha)}-\frac{2i}{g}}{w_{k}^{(\alpha)}-w_{l}^{\alpha}+\frac{2i}{g}}\nonumber,
\end{eqnarray}
where $\alpha=1,2$ reflect the two copies of $\mathfrak{su}(2|2)$
and $S_{0}(p_{k},p_{l})$ is the overall phase of the S-matrix.
\smallskip

Notice that in this construction, the explicit form of the
S-matrix is used. However, since not all bound state S-matrices
are known, one must approach this problem in a different way if
one wishes to find their Bethe equations.

\section{Bethe Ansatz and Yangian Symmetry}

In this section we will generalize the above construction to
arbitrary bound states. We will do this by considering coproducts
of (Yangian) symmetry generators. This formulation allows us to
solve (\ref{eqn;WellDefined}) without knowing the explicit form of
the involved bound state S-matrix. This will lead to the Bethe
equations for arbitrary configurations of bound states.

\subsection{Single excitations}

We will again start by considering a single excitation in the
vacuum
\begin{eqnarray}
|0\rangle = W^{(1)}_{1^{\ell_1}}\ldots W^{(K)}_{1^{\ell_{K}}}.
\end{eqnarray}
As noted above, it suffices to restrict to two bound state
representations. The natural generalization of a single excitation
wave function is:
\begin{eqnarray}\label{eqn;SingleExBound}
|\alpha\rangle:= \sum_{i} \Psi_{i}(y) W^{(1)}_{1^{\ell_1}}\ldots
W^{(i)}_{3^1 1^{\ell_{i}-1}}\ldots W^{(K)}_{1^{\ell_{K}}},\qquad
\Psi_{k}(y) = f(y,p_{k})\prod_{l<k} S^{\mathrm{I,II}}(y,p_{l}),
\end{eqnarray}
Restricted to two sites, the wave function is of the form
\begin{eqnarray}
|\alpha\rangle = f(p_1)W^{(1)}_{\alpha
1^{\ell_{1}-1}}W^{(2)}_{1^{\ell_{2}}} +
f(p_2)S(p_1)W^{(1)}_{1^{\ell_{1}}}W^{(2)}_{\alpha1^{\ell_{2}-1}},
\end{eqnarray}
where we again have chosen a particular normalization.  The
remarkable fact is that one can write this as:
\begin{eqnarray}
\tilde{\Delta}\mathbb{Q}^{1}_{\alpha}|0\rangle:=\left(K_{0}(p_1,p_2)\Delta\mathbb{Q}^{1}_{\alpha}+K_{1}(p_1,p_2)\Delta\hat{\mathbb{Q}}^{1}_{\alpha}\right)|0\rangle,
\end{eqnarray}
with
\begin{eqnarray}
&K_{0}&=
-\sqrt{\frac{2}{g}}\frac{x_2^-\{x_1^-x_2^-(x_1^+x_2^+-1)-x_1^+x_2^+(1+x_1^-[x_1^-+x_1^++x_2^+])\}}{(x_2^-
-x_1^+) (2 x_1^+
   x_2^+x_1^- x_2^--x_1^- x_2^--x_1^+
   x_2^+)}\times\nonumber\\
&&   \times\left[\frac{f(p_2)S(p_1)}{\sqrt{\ell_2} \eta(p_2)}-\frac{e^{-i\frac{p_2}{2}}f(p_1)}{\sqrt{\ell_1} \eta(p_1)}\right]+\frac{e^{-i\frac{p_2}{2}}f(p_1)}{\sqrt{\ell_1}\eta(p_1)}\\
&K_{1}&=\frac{4i\sqrt{2}}{g^{3/2}}\frac{x_1^- x_2^- x_1^+ x_2^+} {
   (x_2^- -x_1^+) (2 x_1^+
   x_2^+x_1^- x_2^--x_1^- x_2^--x_1^+
   x_2^+)}\left[\frac{f(p_2)S(p_1)}{\sqrt{\ell_2} \eta(p_2)}-\frac{e^{-i\frac{p_2}{2}}f(p_1)}{\sqrt{\ell_1}
\eta(p_1)}\right]\nonumber
\end{eqnarray}
For the moment let us keep $f,S$ arbitrary. The invariance of the
S-matrix under Yangian symmetry means that
\begin{eqnarray}
\S\Delta \mathbb{Q}^{1}_{\alpha} = \Delta^{op}
\mathbb{Q}^{1}_{\alpha}\S,\qquad \S\Delta
\hat{\mathbb{Q}}^{1}_{\alpha} =\Delta^{op}
\hat{\mathbb{Q}}^{1}_{\alpha}\S.
\end{eqnarray}
In other words, we find:
\begin{eqnarray}
\S |\alpha\rangle &=& \S
\left(K_{0}(p_1,p_2)\Delta\mathbb{Q}^{1}_{\alpha}+K_{1}(p_1,p_2)\Delta\hat{\mathbb{Q}}^{1}_{\alpha}\right)|0\rangle\nonumber\\
&=&\left(K_{0}(p_1,p_2)\Delta^{op}\mathbb{Q}^{1}_{\alpha}+K_{1}(p_1,p_2)\Delta^{op}\hat{\mathbb{Q}}^{1}_{\alpha}\right)\S|0\rangle\\
&=&\left(K_{0}(p_1,p_2)\Delta^{op}\mathbb{Q}^{1}_{\alpha}+K_{1}(p_1,p_2)\Delta^{op}\hat{\mathbb{Q}}^{1}_{\alpha}\right)|0\rangle,
\end{eqnarray}
since $\S|0\rangle = |0\rangle$. However, we also have
\begin{eqnarray}
|\alpha\rangle_{\pi}&=&\left(K_{0}(p_2,p_1)\Delta^{op}\mathbb{Q}^{1}_{\alpha}+K_{1}(p_2,p_1)\Delta^{op}\hat{\mathbb{Q}}^{1}_{\alpha}\right)|0\rangle.
\end{eqnarray}
This means that (\ref{eqn;WellDefined}) corresponds to requiring
that $K_0$ and $K_1$ are symmetric under interchanging $p_1
\leftrightarrow p_2$. In other words, to find a well-defined
coefficient, we have to solve
\begin{eqnarray}\label{eqn;symmCoeff}
K_{0}(p_1,p_2) = K_{0}(p_2,p_1),\qquad K_{1}(p_1,p_2) =
K_{1}(p_2,p_1),
\end{eqnarray}
for the functions $f$ and $S$.\smallskip

It is straightforward to prove that (\ref{eqn;symmCoeff}) is
equivalent to the equations:
\begin{eqnarray}
K f(p_1) + G f(p_2)S(p_1) &=& f(p_2)\nonumber\\
L f(p_1) + H f(p_2)S(p_1) &=& f(p_1)S(p_2),
\end{eqnarray}
with
\begin{eqnarray}
\begin{array}{lll}
  K=\frac{e^{i\frac{p_1}{2}}}{e^{i\frac{p_2}{2}}}\frac{\sqrt{\ell_2}\eta(p_{2})}{\sqrt{\ell_1}\eta(p_{1})}\frac{x^{+}_{1}-x^{-}_{1}}{x^{+}_{1}-x^{-}_{2}}&\quad & G=
 e^{i\frac{p_1}{2}}\frac{x^{-}_{1}-x^{-}_{2}}{x^{+}_{1}-x^{-}_{2}} \\
  L=e^{-i\frac{p_2}{2}}\frac{x^{+}_{1}-x^{+}_{2}}{x^{+}_{1}-x^{-}_{2}} &\quad& H=
\frac{\sqrt{\ell_1}\eta(p_{1})}{\sqrt{\ell_2}\eta(p_{2})}\frac{x^{+}_{2}-x^{-}_{2}}{x^{+}_{1}-x^{-}_{2}}
\end{array}.
\end{eqnarray}
These equations are solved by the $f,S$ found before, i.e. we
again find (\ref{eqn;SolutionfS}) as unique solution. The
discussion from the previous section holds also here. By fixing
the normalization of the first term to be $f(p_1)$, one finds that
the second term is factorized and fixed.\smallskip

Moreover, notice that from this construction we can read off
elements of the S-matrix. Namely, the coefficients that deal with
the scattering of $W^{(1)}_{\alpha
a^{\ell_1-1}}W^{(2)}_{a^{\ell_2}}$ and
$W^{(1)}_{a^{\ell_1}}W^{(2)}_{\alpha a^{\ell_2-1}}$. We compared
the found coefficients with the explicit known S-matrices and we
indeed find perfect agreement. For example, for $\S^{BB}$ they
coincide with $a_9,a_{10},a_{31}$ and $a_{32}$, cf.
\cite{Arutyunov:2008}.
\smallskip

In conclusion, symmetry of the coefficients uniquely fixes the
form of our wave function. We can now write the wave function,
restricted to two sites, completely in terms of coproducts and as
a consequence (\ref{eqn;WellDefined}) is automatically satisfied.
Finally, the explicit expressions for $K_0,K_1$ are
\begin{eqnarray}
K_{0}(p_{1},p_{2},y)&=&\sqrt{\frac{x_1^-}{x_1^+}}
\sqrt{\frac{x_2^-}{x_2^+}}\frac{y}{(y-x_1^-)(y-x_2^-)}\left[y-\frac{
x_1^-x_2^- x_1^+x_2^+(x_1^-+x_2^-+ x_1^++x_2^+)}{2 x_1^- x_2^- x_1^+ x_2^+ -x_1^- x_2^--x_1^+ x_2^+}\right]\nonumber\\
K_{1}(p_{1},p_{2},y)&=&\frac{4i}{g}\sqrt{\frac{x_1^-}{x_1^+}}
\sqrt{\frac{x_2^-}{x_2^+}}\frac{y}{(y-x_1^-)(y-x_2^-)}\left[\frac{x_1^-x_2^-
x_1^+x_2^+}{2 x_1^- x_2^- x_1^+ x_2^+ -x_1^-
x_2^--x_1^+x_2^+}\right].
\end{eqnarray}
This consideration is valid for \textit{any} bound state numbers
and hence wave function (\ref{eqn;SingleExBound}) is valid for
\textit{any} bound state representations. In particular, all bound
state representations share the same $S^{\mathrm{I,II}}$, and
hence that part of the Bethe equations remains the same.

\subsection{Multiple excitations}

When dealing with two excitations, one needs to introduce a level
$\mathrm{II}$ S-matrix that deals with interchanging $y_1$ and
$y_2$.

\subsubsection*{Fundamental representations}

Let us first restrict to fundamental representations and
reformulate this in terms of coproducts. The wave function was of
the form
\begin{eqnarray}
|\alpha\beta\rangle = |\alpha\beta\rangle_{y_1 y_2} +
\S^{\mathrm{II}}|\alpha\beta\rangle_{y_1 y_2},
\end{eqnarray}
where
\begin{eqnarray}
\S^{\mathrm{II}}|\alpha\beta\rangle_{y_1 y_2} &=&
M(y_1,y_2)|\alpha\beta\rangle_{y_2 y_1} +
N(y_1,y_2)|\beta\alpha\rangle_{y_2 y_1}.
\end{eqnarray}
The natural way to write this would be:
\begin{eqnarray}\label{eqn;TwoExcAnsatz}
|\alpha\beta\rangle_{y_1 y_2} &=&
\left\{(\tilde{\Delta}_{y_{1}}\mathbb{Q}^{1}_{\alpha})(\tilde{\Delta}_{y_{2}}\mathbb{Q}^{1}_{\beta})+
\epsilon_{\alpha\beta}\Delta^{\prime}_{y_1,y_2}\mathbb{L}^{1}_{2}\right\}|0\rangle,
\end{eqnarray}
with
\begin{eqnarray}
\Delta^{\prime}_{y_1,y_2}\mathbb{L}^{1}_{2}&:=&
L_{0}(y_1,y_2,p_1,p_2)\Delta\mathbb{L}^{1}_{2}+L_{1}(y_1,y_2,p_1,p_2)\Delta\hat{\mathbb{L}}^{1}_{2}
\end{eqnarray}
By taking $\alpha=\beta$, one easily checks that the first part is
indeed of the form
$(\tilde{\Delta}_{y_{1}}\mathbb{Q}^{1}_{\alpha})(\tilde{\Delta}_{y_{2}}\mathbb{Q}^{1}_{\beta})$.
Hence, we have to solve $L_0, L_1$ such that our ansatz gives
\begin{eqnarray}
&&\left\{f(y_1,p_1) f(y_2,p_2)S(y_2,p_1) + M f(y_2,p_1) f(y_1,p_2)S(y_1,p_1)\right\} W^{(1)}_{\alpha}W^{(2)}_{\beta}\nonumber\\
&&+  N f(y_2,p_1) f(y_1,p_2)S(y_1,p_1)W^{(1)}_{\beta}W^{(2)}_{\alpha}\nonumber \\
&&+  \epsilon^{\alpha\beta}h(y_1,y_2,p_1)f(y_2,p_1) f(y_1,p_1)W^{(1)}_{2}W^{(2)}_{1}\\
&&+  \epsilon^{\alpha\beta}h(y_1,y_2,p_2)f(y_2,p_2)
f(y_1,p_2)S(y_2,p_1) S(y_1,p_1)W^{(1)}_{1}W^{(2)}_{2}\nonumber,
\end{eqnarray}
where we keep the functions $M,N,h$ arbitrary. This gives two
equations for $L_0$ and two equations for $L_1$. Hence, requiring
symmetry under $p_1\leftrightarrow p_2$ will give us four
equations which can be shown to be equivalent to the following set
of equations:
\begin{eqnarray}\label{eqn;2excSmatrix}
\{f_{12}f_{21}S_{22} + M f_{22}f_{11}S_{12}\} &=&
\{f_{11}f_{22}S_{21} + M f_{21}f_{12}S_{11}\}\frac{D+E}{2}+ N f_{21}f_{12}S_{11}\frac{D-E}{2}\nonumber\\
&&+ \left( -f_{11}f_{21}h_{121} + f_{12}f_{22}S_{11}S_{21}h_{122}
\right)\frac{C}{2}\nonumber\\
 N f_{22}f_{11}S_{12} &=& \{f_{11}f_{22}S_{21} + M f_{21}f_{12}S_{11}\}\frac{D-E}{2}+ N f_{21}f_{12}S_{11}\frac{D+E}{2}\nonumber\\
&&- \left( -f_{11}f_{21}h_{121} + f_{12}f_{22}S_{11}S_{21}h_{122}
\right)\frac{C}{2}.\nonumber\\
f_{11}f_{21}S_{12}S_{22}h_{121} &=& \{f_{11}f_{22}S_{21} + (M-N)
f_{21}f_{12}S_{11}\}\frac{F}{2}\\
&&+f_{11}f_{21}h_{121}\frac{1-B}{2} +
f_{12}f_{22}S_{11}S_{21}h_{122}\frac{1+B}{2}\nonumber\\
f_{12}f_{22}h_{122} &=& -\{f_{11}f_{22}S_{21} + (M-N)
f_{21}f_{12}S_{11}\}\frac{F}{2}\\
&&+f_{11}f_{21}h_{121}\frac{1+B}{2} +
f_{12}f_{22}S_{11}S_{21}h_{122}\frac{1-B}{2},\nonumber
\end{eqnarray}
where, for convenience, we introduced the short-hand notation
$f_{kl}:=f(y_{k},p_{l}), S_{kl}:=S^{\mathrm{II,I}}(y_{k},p_{l}),
M:= M(y_{1},y_{2}),N:= N(y_{1},y_{2})$ and $h_{ijk} :=
h(y_i,y_j,p_k)$. The coefficients $B,C,D,E,F$ are given by
\begin{eqnarray}
B&=&\frac{2 x_1^- x_2^- (x_2^+)^2-(x_1^- x_2^-+1)
   (x_2^-+x_1^+) x_2^++2 x_2^- x_1^+}{(1-x_1^-
   x_2^-) (x_1^+-x_2^-) x_2^+}\nonumber\\
C&=&2 i \eta(p_1) \eta(p_2)\frac{x_2^-}{x_2^+}\frac{
e^{-\frac{ip_1}{2}}(x_2^+-x_1^+)}{(1-x_1^-
   x_2^-)(x_1^+-x_2^-) }\nonumber\\
D&=& \frac{ x_1^--x_2^+}{x_2^--x_1^+
  }\frac{e^{\frac{ip_1}{2}}}{e^{\frac{ip_2}{2}}}\\
E&=&\frac{e^{\frac{ip_1}{2}}}{e^{\frac{ip_2}{2}}}\frac{ (x_1^-
(x_2^- (x_1^--2x_1^+)+1)
   x_1^++(x_1^++x_1^- (x_2^- x_1^+-2))
  x_2^+)}{(1-x_1^- x_2^-)(x_1^+-x_2^-)x_1^+ }\nonumber\\
F&=&2 i
\frac{e^{-\frac{ip_1}{2}}}{\eta(p_1)\eta(p_2)}\frac{(x_1^+-x_1^-)
(x_2^+-x_2^-)
   (x_2^+-x_1^+)}{(1-x_1^-
   x_2^-)(x_1^+-x_2^-)}\nonumber.
\end{eqnarray}
It is readily seen that these expressions coincide with elements
from the fundamental S-matrix. Remarkably, these are exactly the
same equations that arose in the nested Bethe Ansatz. In other
words, the coefficients $B,C,D,E,F$ correspond to elements from
the fundamental S-matrix and we again find
(\ref{eqn;TwoExcSolution}) as the unique solution for
$M,N,h$.\smallskip

It is worthwhile to note that in this way, we have derived the
complete fundamental S-matrix. This derivation differs
fundamentally from the standard derivations
\cite{Arutyunov:2006yd,Beisert:2005tm} since it depends crucially
on the full Yangian symmetry. There might be a relation with
\cite{Torrielli:2008wi} where the fundamental quantum S-matrix was
also derived from Yangian symmetry.\smallskip

To conclude, let us give the explicit solutions for $L_0,L_1$,
\begin{eqnarray}\label{eqn;L0L1}
&L_0& =\frac{ g(y_1-y_2) x^-_1 x^-_2}{2 i (y_1-x^-_1) (y_2-x^-_1)
   (y_1-x^-_2) (y_2-x^-_2)}\times\\
&&\times\left[ (y_1 +
   y_2)-\frac{x^-_1 x^-_2 x^+_1 x^+_2
   (x^-_1+x^-_2+x^+_1+x^+_2)}{2x_1^+ x_2^+x_1^- x_2^--x_1^- x_2^--x_1^+
   x_2^+}-\frac{y_1 y_2 x^+_1 x^+_2
   }{2x_1^+ x_2^+x_1^- x_2^--x_1^- x_2^--x_1^+
   x_2^+}\right\{\nonumber\\
&&(x^+_1+x^+_2-x^-_1-x^-_2) (x^-_1 x^-_2-x^+_1
   x^+_2)
\left.\left.-\left(\frac{1}{x^-_2}+\frac{1}{x^+_1}+\frac{1}{x^+_2}+\frac{1}{x^-_1}\right)
   (x^-_1 x^-_2+x^+_1 x^+_2)\right\}\right]\nonumber\\
   &L_{1}& = \frac{y_1 y_2 x^-_1 x^-_2}{(y_1-x^-_1) (y_2-x^-_1)
(y_1-x^-_2)
  (y_2-x^-_2)} \left[
   (y_1-y_2)+\frac{4 i g^{-1}  x^-_1 x^-_2 x^+_1 x^+_2}{2 x_1^+
   x_2^+x_1^- x_2^--x_1^- x_2^--x_1^+
   x_2^+}\right]\nonumber
\end{eqnarray}
Note that they are indeed manifestly symmetric under
$p_1\leftrightarrow p_2$.

\subsubsection*{Bound states}

One might hope that it is possible to repeat the construction of
previous section and find explicit elements of the bound state
S-matrices again. However, when considering bound states, one
encounters a difficulty. There is a new term, which is of the form
$W^{(i)}_{\alpha\beta1^{\ell_1-2}}$. This term behaves exactly
like $W^{(i)}_{21^{\ell_1-1}}$ and therefore it is hard to rewrite
rewrite everything as in (\ref{eqn;2excSmatrix}) in a unique way
and read of S-matrix elements.\smallskip

Nevertheless, we redo the procedure and try to match
(\ref{eqn;TwoExcAnsatz}) to the obvious generalization of the two
excitation ansatz
\begin{eqnarray}
|\alpha\beta\rangle &=&
\sum_{k<l}\Psi_k(y_1)\Psi_l(y_2)W^{(1)}_{1^{\ell_1}}\ldots
W^{(k)}_{\alpha1^{\ell_k-1}}\ldots
W^{(l)}_{\beta1^{\ell_l-1}}\ldots
W_{1^{\ell_K}}^{(K)}+\nonumber\\
&&+\S^{\mathrm{II}}\sum_{k<l}\Psi_k(y_1)\Psi_l(y_2)W^{(1)}_{1^{\ell_1}}\ldots
W^{(k)}_{\alpha1^{\ell_k-1}}\ldots
W^{(l)}_{\beta1^{\ell_l-1}}\ldots
W_{1^{\ell_K}}^{(K)}+\\
&&+\epsilon^{\alpha\beta}\sum_{k}\Psi_k(y_1)\Psi_k(y_2)h(y_1,y_2,p_k)W^{(1)}_{1^{\ell_1}}\ldots
W^{(k)}_{21^{\ell_k-1}}\ldots W_{1^{\ell_K}}^{(K)}+\nonumber\\
&&+\sum_{k}\Psi_k(y_1)\Psi_k(y_2)g(y_1,y_2,p_k)W^{(1)}_{1^{\ell_1}}\ldots
W^{(k)}_{\alpha\beta1^{\ell_k-2}}\ldots W_{1^{\ell_K}}^{(K)}.
\end{eqnarray}
Or, restricted to two sites
\begin{eqnarray}
|\alpha\beta\rangle &=& \left\{f(y_1,p_1) f(y_2,p_2)S(y_2,p_1) + M f(y_2,p_1) f(y_1,p_2)S(y_1,p_1)\right\} W^{(1)}_{\alpha 1^{\ell_1-1}}W^{(2)}_{\beta 1^{\ell_2-1}}\nonumber\\
&& + N f(y_2,p_1) f(y_1,p_2)S(y_1,p_1)W^{(1)}_{\beta 1^{\ell_1-1}}W^{(2)}_{\alpha 1^{\ell_2-1}} \nonumber\\
&& +g(y_1,y_2,p_1)f(y_1,p_1) f(y_2,p_1)W^{(1)}_{\alpha\beta^{\ell_1-2}}W^{(2)}_{1^{\ell_2}}\nonumber\\
&& +g(y_1,y_2,p_2) f(y_2,p_2)f(y_1,p_2)S(y_1,p_1)S(y_2,p_1)W^{(1)}_{1^{\ell_1}}W^{(2)}_{\alpha\beta^{\ell_2-2}}\\
&& + \epsilon^{\alpha\beta}h(y_1,y_2,p_1)f(y_2,p_1) f(y_1,p_1)W^{(1)}_{21^{\ell_1-1}}W^{(2)}_{1^{\ell_2}}\nonumber\\
&& + \epsilon^{\alpha\beta}h(y_1,y_2,p_2)f(y_2,p_2)
f(y_1,p_2)S(y_2,p_1)
S(y_1,p_1)W^{(1)}_{1^{\ell_1}}W^{(2)}_{21^{\ell_2-1}},\nonumber
\end{eqnarray}
This is indeed possible and imposing symmetry of $L_0,L_1$ as
before provides equations that are uniquely solved by
$g(y_1,y_2,p_k)=\frac{\ell_k-1}{2\ell_k}(1+M-N)$ and
(\ref{eqn;TwoExcSolution}). These factor are, more or less,
expected from fusion. Plugging these solutions back in $L_0,L_1$,
we find the same functions $L_0,L_1$ as in (\ref{eqn;L0L1}) but
one has to bear in mind that now $x^{\pm}$ parameterize bound
state solutions.\smallskip

For completeness, we give the explicit two particle wave function
(restricted to two sites) :
\begin{eqnarray}
|\alpha\beta\rangle &=& \left\{f(y_1,p_1) f(y_2,p_2)S(y_2,p_1) + M f(y_2,p_1) f(y_1,p_2)S(y_1,p_1)\right\} W^{(1)}_{\alpha 1^{\ell_1-1}}W^{(2)}_{\beta 1^{\ell_2-1}}\nonumber\\
&& + N f(y_2,p_1) f(y_1,p_2)S(y_1,p_1)W^{(1)}_{\beta 1^{\ell_1-1}}W^{(2)}_{\alpha 1^{\ell_2-1}} \nonumber\\
&& +\frac{\ell_1-1}{2\ell_1}(1+M-N)f(y_1,p_1) f(y_2,p_1)W^{(1)}_{\alpha\beta^{\ell_1-2}}W^{(2)}_{1^{\ell_2}}\nonumber\\
&& +\frac{\ell_2-1}{2\ell_2}(1+M-N) f(y_1,p_2)f(y_2,p_2)S(y_1,p_1)S(y_2,p_1)W^{(1)}_{1^{\ell_1}}W^{(2)}_{\alpha\beta^{\ell_2-2}}\\
&& + \epsilon^{\alpha\beta}h(y_1,y_2,p_1)f(y_2,p_1) f(y_1,p_1)W^{(1)}_{21^{\ell_1-1}}W^{(2)}_{1^{\ell_2}}\nonumber\\
&& + \epsilon^{\alpha\beta}h(y_1,y_2,p_2)f(y_2,p_2)
f(y_1,p_2)S(y_2,p_1)
S(y_1,p_1)W^{(1)}_{1^{\ell_1}}W^{(2)}_{21^{\ell_2-1}},\nonumber
\end{eqnarray}
with $f,S,h,M,N$ as in
(\ref{eqn;SolutionfS},\ref{eqn;TwoExcSolution}). By construction,
this wave function satisfies (\ref{eqn;WellDefined}) for any bound
state S-matrix. Hence this solves our two excitation case. In
particular one finds that also the level II S-matrix,
$\S^{\mathrm{II}}$ is unchanged for bound states.

\subsection{Bethe equations}

By making use of coproducts and Yangian symmetry, we have found a
way, independent of the explicit form of the S-matrix, to write
down Bethe wave functions. This allowed us to find
$S^{\mathrm{II,I}}$ and we found that the level two S-matrix,
$\S^{\mathrm{II}}$ remains unchanged. Hence all the higher level
factors also remain unchanged. In other words, this yields that
the Bethe equations for any combination of bound states are given
by:
\begin{eqnarray}
e^{ip_{k}L}&=& \prod_{l=1,l\neq k
}^{K^{\mathrm{I}}}\left[S_{0}(p_{k},p_{l})\frac{x_k^+-x_l^-}{x_k^--x_l^+}\sqrt{\frac{x_l^+x_k^-}{x_l^-x_k^+}}\right]^2\prod_{\alpha=1}^{2}\prod_{l=1}^{K_{(\alpha)}^{\mathrm{II}}}\frac{{x_{k}^{-}-y^{(\alpha)}_{l}}}{x_{k}^{+}-y^{(\alpha)}_{l}}\sqrt{\frac{x^+_k}{x^-_k}}\nonumber \\
1&=&\prod_{l=1}^{K^{\mathrm{I}}}\frac{y^{(\alpha)}_{k}-x^{+}_{l}}{y^{(\alpha)}_{k}-x^{-}_{l}}\sqrt{\frac{x^-_k}{x^+_k}}
\prod_{l=1}^{K_{(\alpha)}^{\mathrm{III}}}\frac{y_{k}^{(\alpha)}+\frac{1}{y_{k}^{(\alpha)}}-w_{l}^{(\alpha)}+\frac{i}{g}}{y_{k}^{(\alpha)}+\frac{1}{y_{k}^{(\alpha)}}-w_{l}^{(\alpha)}-\frac{i}{g}}\\
1&=&\prod_{l=1}^{K_{(\alpha)}^{\mathrm{II}}}\frac{w_{k}^{(\alpha)}-y_{k}^{(\alpha)}-\frac{1}{y_{k}^{(\alpha)}}+\frac{i}{g}}{w_{k}^{(\alpha)}-y_{k}^{(\alpha)}-\frac{1}{y_{k}^{(\alpha)}}-\frac{i}{g}}
\prod_{l\neq
k}^{K_{(\alpha)}^{\mathrm{III}}}\frac{w_{k}^{(\alpha)}-w_{l}^{(\alpha)}-\frac{2i}{g}}{w_{k}^{(\alpha)}-w_{l}^{\alpha}+\frac{2i}{g}}\nonumber,
\end{eqnarray}
with
\begin{eqnarray}
x_{k}^{+}+\frac{1}{x_{k}^+}-x_{k}^{-}-\frac{1}{x_{k}^-} = \frac{2i
\ell_{k}}{g},\qquad \frac{x_{k}^+}{x_{k}^-} = e^{ip_k}.
\end{eqnarray}
However, note that apart from the parameters $x^{\pm}$, the phase
factor $S_{0}(p_{k},p_{l})$ also implicitly depends on the
considered bound states via \cite{Arutyunov:2008}:
\begin{eqnarray}\label{eqn;FullPhase}
S_{0}(p_{1},p_{2})&=&\left(\frac{x_{1}^{-}}{x_{1}^{+}}\right)^{\frac{\ell_2}{2}}\left(\frac{x_{2}^{+}}{x_{2}^{-}}\right)^{\frac{\ell_1}{2}}\sigma(x_{1},x_{2})\times\nonumber\\
&&\times\sqrt{G(\ell_2-\ell_1)G(\ell_2+\ell_1)}\prod_{l=1}^{\ell_1-1}G(\ell_2-\ell_1+2l).
\end{eqnarray}
The derived Bethe equations coincide with the equations one
expects from a fusion procedure. This justifies a fusion procedure
at the level of the Bethe ansatz equations.\smallskip

Finally, we will present the Bethe equations in the mirror theory
since they will have applications there for the TBA program. In
the mirror theory the parameters $a,b,c,d$ describing the symmetry
algebra have the same dependence on $x^{\pm}$ as in the original
theory. However $x^{\pm}$ are now dependent on the mirror momentum
\begin{eqnarray}\label{eqn;xpmMirror}
x_i^{\pm}(\tilde{p})= \frac{\ell_i}{2g}\left(-\sqrt{1+\frac{4g
^2}{\ell_i^2+\tilde{p}^2}}\mp1
\right)\left(-\frac{\tilde{p}}{\ell_i}-i\right).
\end{eqnarray}
To analyze bound states in the mirror theory it is more convenient
to pick up a vacuum build up out of fermions ($\mathfrak{sl}(2)$
sector). The derivation can be repeated for this case and the
Bethe equations are given by \cite{Arutyunov:2007tc}
\begin{eqnarray}
e^{i\tilde{p}_{k}L}&=& \prod_{l=1,l\neq k
}^{K^{\mathrm{I}}}S_{0}(\tilde{p}_{k},\tilde{p}_{l})^{2}\prod_{\alpha=1}^{2}\prod_{l=1}^{K_{(\alpha)}^{\mathrm{II}}}\frac{{x_{k}^{+}-y^{(\alpha)}_{l}}}{x_{k}^{-}-y^{(\alpha)}_{l}}\sqrt{\frac{x^-_k}{x^+_k}}\nonumber \\
-1&=&\prod_{l=1}^{K^{\mathrm{I}}}\frac{y^{(\alpha)}_{k}-x^{+}_{l}}{y^{(\alpha)}_{k}-x^{-}_{l}}\sqrt{\frac{x^-_k}{x^+_k}}
\prod_{l=1}^{K_{(\alpha)}^{\mathrm{III}}}\frac{y_{k}^{(\alpha)}+\frac{1}{y_{k}^{(\alpha)}}-w_{l}^{(\alpha)}+\frac{i}{g}}{y_{k}^{(\alpha)}+\frac{1}{y_{k}^{(\alpha)}}-w_{l}^{(\alpha)}-\frac{i}{g}}\\
1&=&\prod_{l=1}^{K_{(\alpha)}^{\mathrm{II}}}\frac{w_{k}^{(\alpha)}-y_{k}^{(\alpha)}-\frac{1}{y_{k}^{(\alpha)}}+\frac{i}{g}}{w_{k}^{(\alpha)}-y_{k}^{(\alpha)}-\frac{1}{y_{k}^{(\alpha)}}-\frac{i}{g}}
\prod_{l\neq
k}^{K_{(\alpha)}^{\mathrm{III}}}\frac{w_{k}^{(\alpha)}-w_{l}^{(\alpha)}-\frac{2i}{g}}{w_{k}^{(\alpha)}-w_{l}^{\alpha}+\frac{2i}{g}}\nonumber,
\end{eqnarray}
where $\tilde{p}$ is the mirror momentum and $x^{\pm}$ are now
given by (\ref{eqn;xpmMirror}). The $-1$ on the left hand side is
due to anti-periodic boundary conditions for the fermions
\cite{Arutyunov:2007tc}. For a detailed account of the mirror
theory and its analytic properties we refer to
\cite{Arutyunov:2007tc}.

\section*{Acknowledgements}

I am indebted to G. Arutyunov and S. Frolov for many valuable
discussions. I would also like to thank A. Torrielli for comments
on the manuscript. This work was supported in part by the EU-RTN
network \textit{Constituents, Fundamental Forces and Symmetries of
the Universe} (MRTN-CT-2004-005104), by the INTAS contract
03-51-6346 and by the NWO grant 047017015.

\bibliographystyle{JHEP}
\bibliography{LitRmat}

\end{document}